\documentclass[12pt]{JHEP}
\usepackage[dvips]{epsfig}
\usepackage{epsfig}
\usepackage{graphicx}

\title{Axisymmetric metrics in arbitrary dimensions}
\author{Christos Charmousis\\
LPT\footnote{Unit\'e Mixte de Recherche du CNRS (UMR 8627).}, 
Universit\'e de Paris-Sud, B\^at 210, 91405 Orsay CEDEX, France}
\author{Ruth Gregory\\
Centre for Particle Theory, University of Durham\\
South Road, Durham, DH1 3LE, U.K.}

\abstract{
We consider axially symmetric static metrics in arbitrary dimension,
both with and without a cosmological constant. The most obvious such
solutions have an $SO(n)$ group of Killing vectors representing
the axial symmetry, although one can also consider abelian groups
which represent a flat `internal space'. We
relate such metrics to lower dimensional dilatonic cosmological metrics
with a Liouville potential. We also develop a duality relation between
vacuum solutions with internal curvature and those with
zero internal curvature but a cosmological constant.
This duality relation gives a solution generating
technique permitting the mapping of different spacetimes.
We give a large class of solutions to the vacuum or cosmological
constant spacetimes. We comment on the extension of the C-metric 
to higher dimensions and provide a novel solution for a braneworld
black hole. }
\preprint{gr-qc/0306069\\
LPT-ORSAY/0341\\
DCPT-03/68\\IPPP-03/34}

\def\half{\textstyle{1\over2}}

\def\quarter{\textstyle{1\over4}}
\def\sech{\,{\rm sech}\,}
\def\ie{{\it i.e.,}}

\newcommand{\be}{\begin{equation}}
\newcommand{\ee}{\end{equation}}
\newcommand{\bea}{\begin{eqnarray}}
\newcommand{\eea}{\end{eqnarray}}
\newcommand{\bml}{\begin{mathletters}}
\newcommand{\eml}{\end{mathletters}}

\begin{document}

\section{Overview}

Most systems of physical interest in nature exhibit a certain amount
of symmetry: a star is roughly spherically symmetric, a galaxy roughly
axisymmetric, our universe has roughly constant spatial curvature.
These three examples share the feature that the dominant interaction
governing their large scale structure is gravity.
Einstein's theory of general relativity in principle describes a
highly nonlinear interaction, yet, if one applies a few
physically motivated coordinate choices reflecting the symmetry of
the system, the gravitational equations become easier,
and sometimes straightforward, to solve. Indeed, in four
dimensions we have a huge range of known solutions covering a wide
selection of physically relevant situations \cite{exact}.

Four dimensions are of course very special for gravity. It is the smallest
spacetime dimension in which general relativity becomes nontrivial, in the
sense that gravity can propagate through spacetime and objects interact
with one another -- it is when the nonlinearities of gravity
truly start to show up. Yet four dimensions are also a suitably low enough
number that we have a pretty good handle on gravitational interactions,
and can in fact construct an encyclopaedia \cite{exact} of exact solutions
rich enough that we can find metrics for all but the most complicated
physical systems.

It seems however that string theory is compelling us
to have more than four -- in fact eleven -- dimensions.
While the true nature of this eleven dimensional
M-theory has yet to be elucidated, there is likely to be some
energy range in which physics is described well by a classical
(super)gravity field theory in more than four dimensions.
As a result, an increasingly important role has emerged for
the study of supergravity solutions in 10 and 11 dimensions.
Indeed, the celebrated adS/CFT
correspondence \cite{ADSCFT} arises from consideration of the
consistency of stringy versus supergravity pictures of a D-brane.

Furthermore in the last few years, and largely
motivated by string theory, there has been a huge interest
in toy models where our universe is a submanifold,
or {\it braneworld}, embedded in a higher dimensional
spacetime \cite{LOSW,ADD,RS1,RS2}. Many interesting
scenarios and ideas have been put forward such as the one by Randall and
Sundrum \cite{RS2}, where the ``extra'' fifth dimension, can be infinite
provided that the bulk spacetime is negatively curved.
Of course, given that these models allegedly describe our universe,
it is crucial to study their cosmology, as well as other strongly
gravitating phenomena such as black holes. Brane cosmology of 
codimension one
(\ie\ where there is only one extra dimension) is well understood
\cite{NONC,BCOS,BCG}, and cosmological perturbations
of such models have been studied \cite{BPT}, although there
still remain many important unanswered questions.

On the other hand, progress in finding the five dimensional solution
to a black hole which is localised
on the four dimensional brane-Universe has proved much
more elusive. While one can describe a black hole on the brane
by simply extending the Schwarzschild solution into the bulk \cite{CHR},
this solution is singular at the adS horizon, and in addition is
unstable \cite{BSINS}. Clearly the physically correct solution
will be localized in the bulk, and should correspond to a correction
or extension of a Schwarzschild
black hole in a four dimensional Universe much like the
standard FLRW equation of cosmology
is extended in the cosmological version of the RS model \cite{NONC}.
Progress in this direction has been realised numerically
\cite{BHNUM}, but it would clearly be preferable to have an
analytic solution to the problem if at all possible.

An interesting indirect investigation by Emparan, Horowitz
and Myers \cite{EHM} gave a lower dimensional solution to
the problem, namely a three-dimensional black hole in a three-dimensional
braneworld living in four-dimensional adS spacetime. They realised
that the bulk metric should describe an accelerating black hole,
the C-metric \cite{cmetric} (for a full reference list see
\cite{exact}-for the adS case more recently see \cite{lemos}).
To realise why this is so, one must remember that a domain
wall, \cite{VIS}, generically has a Rindler horizon \ie\ it has an
accelerating trajectory in an otherwise constant curvature
spacetime. Any black hole residing on the wall (or braneworld universe)
must also accelerate by the same amount to ``keep up'' with the
wall's motion.  Unfortunately, to date it has been
impossible to find the generalisation of the C-metric in higher than four
dimensions. Recently Emparan, Fabbri and Kaloper \cite{EFK}
(see also \cite{Tanaka})
put forward a conjecture relating the above problem to the
adS/CFT correspondence. A classical bulk solution
describing a black hole localised on the brane corresponds to a
quantum-corrected black hole for the four-dimensional observer living on
the boundary-brane. If this conjecture is true it is not at all
surprising that we have difficulty in finding
the extension of the C-metric. 

The C-metric in four dimensions
belongs to the so called Weyl class of metrics: static and axisymmetric
solutions of Einstein's equations.
General Relativity in more than four dimensions is rather more
rich and complicated. Weyl metrics, which we shall be discussing here,
correspond to an integrable system in four dimensions
but not in higher dimensions.
Furthermore, in four dimensions event horizons are topologically spherical,
but in higher dimensions we can have solutions with not only
spherical (for early work in the context of string theory see \cite{MP}),
but also hyper-cylindrical (for the canonical string theoretic $p$-brane
solutions see \cite{HS}), and even genuinely toroidal event
horizons \cite{ER2}, although it would seem
that many of the more exotic topologies have instabilities \cite{GL}.
However, the story does not stop there, in fact, a
consideration of the endpoint of these instabilities
leads us to suspect that there might even be ripply
horizons \cite{ripple} (first pointed out in \cite{GLE}),
as well as the more obvious
possibility of periodic black hole solutions \cite{BHNUM}.

While we have an excellent catalogue of solutions in four dimensions, the
situation in higher dimensions is more patchy. The known solutions are by
and large mostly one-dimensional, in the sense that the metric depends
on only one coordinate.
In the case of the Horowitz-Strominger black branes
\cite{HS}, or the cosmic $p$-branes \cite{CPB}, (a set of
Poincar\'e boost-symmetric solutions corresponding to pure
gravitating brane-like sources), the solutions depend on
a single coordinate transverse to the brane.
While it is true that most physical systems
can indeed be reduced to effectively depend on
only one variable (such as time in the case of cosmology, or
`radius' in the case of a static star) in practise, we also
want to have a system which depends on two variables - such as
an axially symmetric mass distribution, like a galaxy,
an anisotropic cosmology or (more recently) a braneworld metric.
The known ``two-dimensional'' solutions tend to
have some special symmetry which allows their integrability, isotropy
in the case of cosmological braneworld solutions \cite{BCG},
or supersymmetry in the case of `intersecting' brane solutions \cite{intsct}
-- although as these solutions tend to be delocalized they are not as
intuitive as might first appear. Genuinely two-dimensional solutions, such
as would represent a higher dimensional galaxy, are much harder to find.
First steps in this direction are the work of Emparan and Reall \cite{er},
who analyse the case where the Killing vectors representing the symmetries
of the spacetime commute, and also the work of reference \cite{christos}, 
where special classes of dependence were considered.

In this paper, we are interested in the general axisymmetric metric,
which we will call a Weyl metric, after the work of Weyl \cite{Weyl}
on four dimensional static axisymmetric metrics.
(Note that by analytic continuation, we could also consider
metrics depending on time and one of the spatial coordinates,
which we will call cosmological or Thorne metrics after the work
of Thorne \cite{Thorne} on cylindrically symmetric metrics.)
The fact that the metric depends on two coordinates is not itself
a guarantee that the spacetime is genuinely {\it two dimensional},
as the situation with braneworlds so eloquently
illustrates \cite{BCG,GP}. Rather, it is the fact that the `transverse'
space, or that part of the metric which does not depend on the two
main coordinates, is split into (at least) two separate subspaces
spanned by mutually commuting sets of Killing vectors.
Each such subspace introduces an additional degree of freedom
into the metric -- a breathing mode or modulus -- and if one of these
subspaces does not have zero curvature (Emparan and Reall \cite{er}
dealt with the zero curvature case) then, roughly speaking, this
curvature introduces a source term into the equation of motion
for that particular breathing mode.

To see why this is problematic,
it is useful to think of this multidimensional spacetime as an
effective two-dimensional field theory, by simply dimensionally
reducing over the nonessential coordinates. In this case, we get
a scalar field which represents the two dimensional part of the
metric, and each breathing mode introduces an additional scalar
into this two-dimensional theory. If all we have are kinetic
terms, then this is an integrable
system, and we can prescriptively solve the general metric.
If however there is a curvature term present, then this has the
effect of introducing a Liouville potential into the theory,
which renders the problem a great deal more subtle and complex.

What we try to do in this paper is to determine to what extent
we can give a prescription for solving the general axisymmetric
problem. We find a transformation which reduces the Einstein
equations to a relatively simple form and give a set of methods
for their solution, remarking on the level of completeness of
our prescription. We describe a duality relation which relates vacuum
spacetimes with `internal' curvature and cosmological constant 
spacetimes without this `internal' curvature. 
We present a large range of solutions, some already
discovered, and some new ones. The layout of the paper is as follows:
after first reviewing the four-dimensional case for reference,
we examine the general axisymmetric metric with curvature, and
present our reduction of the Einstein equations into two canonical
`dual' forms. In the following sections we treat the cases
of zero curvature and cosmological constant (the Emparan-Reall
case) presenting some extensions of four-dimensional solutions,
and remarking on the intuition these simpler metrics give. We then
discuss vacuum spacetimes with internal curvature, and cosmological 
constant spacetimes without internal curvature, finding a new range of
solutions. Finally, we apply our results to braneworld black holes
and the higher dimensional C-metric. We present a linearized
gravity prescription for general sources, and a new regular 
exact solution for black holes on adS branes.

\section{Four dimensional Weyl metrics: A review}

In four dimensions, time and axisymmetry mean that the metric
depends only on two remaining coordinates, $r$ and $z$ say, and we
can write the metric into the block diagonal form
\be
ds^2 = e^{2\lambda} dt^2 - e^{2(\nu-\lambda)} (dr^2 + dz^2)
- \alpha^2 e^{-2\lambda} d\phi^2 \label{4dweylmet}
\ee
called the Weyl canonical form.
Note that although we have three metric functions, the transverse $t$
and $\phi$ spaces are both intrinsically Ricci flat (somewhat trivially),
which turns out to be crucial in the integrability of the
Einstein equations:
\bea
\Delta\alpha &=& - \alpha e^{2(\nu-\lambda)} \left [ T^r_r + T^z_z \right ]
\label{4dalph} \\
\Delta \lambda + {\nabla \lambda \cdot \nabla \alpha \over \alpha} &=&
{\half} e^{2(\nu-\lambda)} \left [
T^t_t - T^r_r - T^z_z - T^\phi_\phi \right ] \label{4dlam} \\
\Delta\nu + (\nabla\lambda)^2 &=& - e^{2(\nu-\lambda)} T^\phi_\phi
\label{4nu} \\
{\partial_\pm^2\alpha\over\alpha} + 2 (\partial_\pm\lambda)^2
- 2 \partial_\pm \nu {\partial_\pm\alpha\over\alpha} &=&
T_{rr} - T_{zz} \pm 2iT_{rz} \label{4dc}
\eea
where $T_a^b$ is the energy momentum tensor (with the factor $8\pi G$
absorbed), $\Delta$ is the two dimensional Laplacian $(\partial_r^2
+ \partial_z^2 = \partial_+\partial_-)$, with $\partial_\pm
= \partial_r \mp i\partial_z$ the derivatives with respect
to the complex coordinates $\zeta = (r+iz)/2$ and $\bar\zeta$.

In the absence of matter or a cosmological constant, these have a very elegant
solution: one simply fixes the conformal gauge freedom remaining in
the metric (\ref{4dweylmet}) by setting $\alpha \equiv r$, which
is consistent with (\ref{4dalph}). This then means (\ref{4dlam})
becomes a cylindrical Laplace equation for $\lambda$, with solution
\be
\lambda = -{1\over 4\pi} \int {S({\bf r}') d^3 {\bf r}'\over |{\bf r}
- {\bf r}'|} \label{4lamsol}
\ee
for a source with energy density $S({\bf r})$. Note then that the metric
component $\lambda$, is nothing but the Newtonian source of
axial symmetry. In turn
$\nu$ is determined
from $\lambda$ via (\ref{4dc}).
Since the $\lambda$ equation is linear, its solutions
can obviously be superposed -- the nonlinear nature of
Einstein gravity showing up in the solution of $\nu$.
Note that as regularity
of the $r$-axis requires $\nu(0,z)=0$, in general there will
be conical singularities when regular solutions are superposed. These
can be interpreted as strings or struts supporting the static sources
in equilibrium. Let us now describe briefly some solutions of physical
interest which we would like to be able to reproduce in more than
four dimensions.

\subsection{Black hole spacetimes}

Physical solutions of particular interest are of course black hole
and multi-black hole solutions, which correspond to line mass sources
in (\ref{4lamsol}) for the Newtonian picture
(for a clear and concise description see \cite{Fay}) -- a semi-infinite
line mass source actually corresponding to an acceleration horizon.

To see this, input into (\ref{4lamsol}) a line source with
unit mass per unit length,
$S({\bf r}) = \delta(r)/r$ for $z\in[c_3,c_4]$:
\be
\lambda_S = -{1\over2} \int_{c_3}^{c_4} {dz'\over[r^2 + (z-z')^2]^{1/2}}
= {\half} \ln {R_3 - z_3 \over R_4 - z_4}
= {\half} \ln {X_3 \over X_4} \label{Schlam}
\ee
where
\be
z_i = z - c_i \qquad , \qquad R_i^2 = r^2 + z_i^2
\ee
Integration of (\ref{4dc}) then gives
\be
\nu_S = {1\over2} \ln {(R_3R_4 + z_3z_4 + r^2) \over 2R_3R_4}
= {1\over2} \ln {Y_{34} \over 2R_3R_4}
\label{Schnu}
\ee
Although this does not appear to be a Schwarzschild black hole,
defining $M = c_4-c_3$, the simple transformation
\be
z=(\rho-M)\cos\theta \;\; , \;\;\; r^2 = \rho(\rho-2M) \sin^2\theta
\label{w2sch}
\ee
in fact returns the metric to its standard spherical form.

Now we can consider superposing solutions for $\lambda$, to
build up multi-black hole solutions in the way first described by
Israel and Khan \cite{IK}. For the simplest case of adding a second
black hole to (\ref{Schlam}) of mass $M' = c_2-c_1$
(with $c_1<c_2<c_3<c_4$) we have:
\bea
\lambda &=& {1\over2} \ln {X_1 X_3\over X_2 X_4}\label{iklam}\\
\nu &=& {1\over2} \ln { Y_{21} Y_{34} Y_{23} Y_{41} \over
4 R_1R_2R_3R_4Y_{13} Y_{24} } + \nu_0 \label{iknu}
\eea
where $Y_{ij}$ was defined implicitly in (\ref{Schnu}).

The reason for the constant $\nu_0$ is that for regularity of the
$z$-axis, we require $\nu(0,z)=0$, as already mentioned. For a
single rod potential, the Schwarzschild black hole, we can make
the axis everywhere nonsingular except for the event horizon,
$z\in [c_3,c_4]$. However, once we add another mass source, due to their
mutual attraction,
we can no longer have a static spacetime {\it and} a nonsingular
$z$-axis. Calculating $\nu$ on the axis gives
\be
\nu(0,z) = \cases{\nu_0 & $z>c_4$ or $z<c_1$\cr
\ln{(c_3-c_2)(c_4-c_1)\over(c_3-c_1)(c_4-c_2)} +\nu_0 <  \nu_0& $c_2<z<c_3$.}
\ee
Thus if $\nu_0=0$, meaning that the $z$-axis is regular away
from the black holes, then
inbetween the two black holes $\nu(0,z)<0$, hence
we have a conical excess or strut, which can be interpreted as
supporting the black holes in their static equilibrium. Similarly, if
we choose $\nu_0$ to have a regular axis inbetween the black holes, we
have a conical deficit extending from each black hole out to infinity,
which can be interpreted as a string suspending the black holes in
their (unstable) equilibrium.

\subsection{The Rindler and C-metrics}\label{subsec:Crin}

Now consider a semi-infinite line mass (SILM) which can formally be
obtained from (\ref{Schlam}) by letting $c_3, c_4 \to \infty$ with
$c_3/c_4 = {\hat c}_3$ a constant. Then
simultaneously rescaling the dimensionful
coordinates ${\hat t} = t /(2c_4A)$, ${\hat r} = 2c_4Ar$,
${\hat z} = 2c_4Az$, gives the Rindler metric:
\be
ds^2 = A{\hat X}_3 d{\hat t}^2 - {1\over 2A{\hat R}_3} \left (
d{\hat r}^2 + d{\hat z}^2 \right ) - {{\hat r}^2\over A{\hat X}_3} d\phi^2
\label{canrnw}
\ee
One can put this in a more transparent `C-metric' form with the
coordinate transformation
\be
{\hat r} = {\sqrt{1-x^2} \sqrt{y^2-1}\over A(x+y)^2} \;\;\;\;,\;\;\;
{\hat z} ={1+xy\over A(x+y)^2}
\ee
where $A=1/(2{\hat c}_3)$ is the acceleration parameter for the
Rindler metric, and which gives
\be
ds^2 = {1\over (x+y)^2} \left [ (y^2-1) A^{-2} dt^2
- {dy^2\over (y^2-1)}
-{dx^2\over(1-x^2)} - (1-x^2)d\phi^2 \right ]
\label{rinc}
\ee
To get flat space use
\bea\label{rintr}
T &=& {\sqrt{y^2-1}\over A(x+y)} \sinh At \;\;\;\;\;\;\;\;
Y = {\sqrt{1-x^2}\over A(x+y)} \cos\phi \nonumber\\
X &=& {\sqrt{y^2-1}\over A(x+y)} \cosh At \;\;\;\;\;\;\;\;
Z = {\sqrt{1-x^2}\over A(x+y)} \sin\phi
\eea
Thus we see how flat space can be written in a variety of
forms, the `C-metric' form (\ref{rinc}) being the most obvious
way of tailoring a coordinate system to an accelerating observer.
On the other hand the canonical Weyl form (\ref{canrnw}) allows us
to make contact with Newtonian potential theory and building up
spacetimes with acceleration horizons and black holes - the
simplest of course being the C-metric \cite{cmetric} which we turn to next.

As with the Rindler metric, if we take $c_4 \to \infty$, and rescale
$t,r,z$ as before, then we get the C-metric in Weyl coordinates:
\be
ds^2 = {{\hat X}_1{\hat X}_3\over A{\hat X}_2} d{\hat t}^2
- {e^{2\nu_0}{\hat Y}_{12}{\hat Y}_{23}\over 4AR_1R_2R_3 {\hat Y}_{13}}
\left ( d{\hat r}^2 + d{\hat z}^2 \right )
- {{\hat r}^2 {\hat X}_2\over A{\hat X}_1{\hat X}_3} d\phi^2
\ee
The C-metric has
the canonical form (see also more recently \cite{Hong:2003gx}),
\be
\label{cmetric}
ds^2 = A^{-2}(x+y)^{-2}[{F(y) dt^2 - F^{-1}(y) dy^2 - G(x)
d\phi^2 - G^{-1}(x) dx^2]},
\ee
where
\be
\label{cmetric1}
G(x) = 1-x^2-2mAx^3\;, \;\;\; F(y) = -1+y^2 - 2mAy^3
\ee
Here, $m$ represents the mass of the black holes, and $A$ their
acceleration. In the flat space limit, $A^{-1}$ represents
half the distance of closest approach.  Let us write
$x_1<x_2<x_3$ for the roots of $G$. Then, in order to obtain 
the correct signature, we must have $x_2<x<x_3$ and
$-x_2<y<-x_1$. The coordinates cover only one patch of the full spacetime 
corresponding to the exterior spacetime of one accelerating hole up 
to its acceleration horizon, which is located at $y=-x_2$. 
The coordinate singularity at $y = -x_1$ corresponds to the 
event horizon of the black hole, whereas the black hole singularity
itself is
screened and resides at $y\rightarrow \infty$. 
The conical deficit sits along $x=x_2$, 
while $x=x_3$ points towards the other black hole,
which means that $\phi$ has periodicity $4\pi/|G'(x_3)|$.
We will return to the C-metric in the conclusions.

\section{General set-up in arbitrary dimensions and the equivalence}

In more than four dimensions, an axisymmetric solution will in general
have a nonabelian group of Killing symmetries (the abelian case was
formally analyzed in \cite{er}), usually corresponding to SO($n$) 
symmetry for some $n$. 
It turns out that this has a significant effect on the solubility
of the Einstein equations, as the metric now contains closed Killing
surfaces of constant curvature, which give rise to source terms
in the Einstein equations somewhat analogous to cosmological terms
in the four-dimensional equations (\ref{4dalph}-\ref{4dc}).
This is what we mean by {\it internal curvature}.
To see this explicitly, consider the Weyl metric
corresponding to an axisymmetric $p$-brane living in $D=p+n+3$ dimensions:
\be
ds^2 = A^2 \left [ dt^2 - dy_p^2 \right]
-B^2 (dr^2 + dz^2 ) - C^2 dx_{n,\kappa}^2
\label{genwmet}
\ee
This represents a Poincare $p$-brane with $(n+2)$-orthogonal directions, 
and the $x$-space has curvature $\kappa=0,-1,1$.
In essence, the $p$-spatial directions of the brane are superfluous, and we
are primarily interested in $p=0$, but for generality, we will maintain
the parameter $p$ while deriving the equations of motion and many of the
solutions.

In seeking the equivalent of the Weyl canonical form we first calculate
the Ricci curvature of  this metric:
\bea
R^t_t &=& {1\over B^2} \left [ {\Delta A\over A} + p{(\nabla A)^2\over A^2}
+ n {\nabla A \cdot \nabla C \over AC} \right ] \label{Riccit}\\
R^x_x &=& {1\over B^2} \left [ {\Delta C\over C} + (n-1){(\nabla C)^2\over C^2}
+ (p+1) {\nabla A \cdot \nabla C \over AC} \right ] - {(n-1)\kappa \over C^2}
\label{Riccix} \\
R^z_z &=& {p+1\over B^2} \left [ {{\ddot A}\over A} + {A'B'\over AB}
- {{\dot A} {\dot B}\over AB} \right ]
+ {n\over B^2} \left[ {{\ddot C}\over C} +
{C'B'\over CB} - {{\dot C} {\dot B} \over CB} \right ] +
{\Delta \ln B\over B^2} \label{Ricciz} \\
R^r_r &=& {p+1\over B^2} \left [ {A''\over A} - {A'B'\over AB}
+ {{\dot A} {\dot B}\over AB} \right ] + {n\over B^2} \left[ {C''\over C}
- {C'B'\over CB} + {{\dot C} {\dot B} \over CB} \right ] +
{\Delta \ln B\over B^2} \label{Riccir} \\
R_{rz} &=& - n {{\dot C}'\over C} - (p+1){{\dot A}'\over A}
+ n {C'{\dot B} \over CB} + n {{\dot C}B'\over CB}
+ (p+1) {A'{\dot B} \over AB} + (p+1) {{\dot A}B'\over AB} \label{Riccirz}
\eea
where a prime denotes $\partial/\partial r$, and a 
dot $\partial/\partial z$.
It is immediately apparent that the appropriate generalisation of the
$\alpha$ variable from the four-dimensional case is
\be
\label{alphadefn}
\alpha = A^{p+1} C^n,
\ee
however, we have some freedom in how we spread $\alpha$ across $A$ and $C$.
To see why this might be relevant, consider first writing
\bea
\ln A &=& a\phi \label{Aphi} \\
\ln B &=& \chi - {\gamma\over 2} \phi - {n-1\over 2n} \ln \alpha \label{Bchi}
\eea
where we have set
\be
\label{podi}
a=\pm\sqrt{{n\over 2(p+1)(n+p+1)}},\qquad
\gamma=\pm\sqrt{{2(p+1)\over n(n+p+1)}}
\ee
In other words, we have rewritten the metric as:
\be
\label{metric}
ds^2 = e^{2a\phi} \left [ dt^2 - dy_p^2 \right]
- e^{-\gamma \phi} \left \{
\alpha^{-(n-1)/n} e^{2\chi} (dr^2 + dz^2 )
+ \alpha^{2/n} dx_{n,\kappa}^2 \right \}
\ee
By rewriting the metric specifically in this way, the connection
to cosmological dilatonic metrics can be made apparent, since upon
``dimensional reduction'' (ignoring the space or time-like nature of the
dimensions being reduced over) over $t$ and $y$ we obtain a 
cosmological-type
metric in an $(n+2)$-dimensional spacetime with a Liouville potential
if the cosmological constant in $D$-dimensions does not vanish:
\be
\label{Dact}
S_D = \int d^D X \left [ -R_D + 2 \Lambda \right] \propto
\int d^{n+2}x \left [ -R_{n+2} + {1\over2} (\nabla\phi)^2
+ 2 \Lambda e^{-\gamma\phi} \right]
\ee
Note that $\gamma$ in (\ref{podi}) relates to the exponential Liouville
coupling in (\ref{Dact}).
Of course, strictly speaking to get a metric with a cosmological interpretation
we have to double-analytically continue $t \to i\chi$ and $r\to i\tau$,
however the overall symmetries of the equations of motion remain
essentially the same, simply swapping operators from Laplacian to Dalembertian.
Such metrics arise in many string-inspired cosmological models (see
for example \cite{SBW}, \cite{BWS}), \cite{Langlois:2001dy}.

Using (\ref{Aphi}-\ref{Bchi}), the equations of motion take the form
\bea
\Delta \alpha &=& -2\Lambda \alpha B^2 + n(n-1)\kappa \alpha G\label{alpheq} \\
a\left ( \Delta \phi + \nabla \phi \cdot {\nabla\alpha\over\alpha} \right)
&=& -{2\Lambda B^2\over(D-2)} \label{phieq} \\
\Delta \chi + {\quarter} (\nabla \phi)^2 &=& - {\Lambda B^2\over n}
-{(n-1)\kappa G\over2}\label{chieq} \\
{\partial_\pm^2\alpha\over\alpha} + {\half} (\partial_\pm\phi)^2
- 2 \partial_\pm \chi {\partial_\pm\alpha\over\alpha} &=& 0 \label{consteq}
\eea
where as before $2\partial_\pm = \partial/\partial(r\pm iz)$, and
\bea
B^2 &=& e^{2\chi} \alpha^{-(n-1)/n} e^{-\gamma\phi} \label{Bform}\\
G &=& B^2/C^2 = e^{2\chi} \alpha^{-(n+1)/n}\label{Gform}
\eea
The operator expressions appearing on the left hand side of
(\ref{alpheq}-\ref{consteq}) are independent
of dimension unlike the expressions, (\ref{Riccit}-\ref{Ricciz}).
Therefore upon switching-off the curvature scales
$\kappa$ and $\Lambda$ we immediately obtain the equivalent of the Weyl
canonical form in arbitrary dimensions.
Note also that if $\Lambda = 0$, then (\ref{phieq}) states that
$\alpha \nabla\phi$ is divergence
free in the $(r,z)$ plane.

The writing of the field equations in a canonical form is not unique, we
could instead choose to write
\bea
\ln A &=& {\bar a}{\bar\phi}
+ {1\over {\bar D}-2}\ln \alpha \label{Aphidual} \\
\ln B &=& {\bar\chi} - {{\bar D}-3\over 2({\bar D}-2)}
\ln \alpha \label{Bchidual}
\eea
for a $\bar D$-dimensional spacetime.
The metric then takes quite a different form from (\ref{metric}) and reads,
\be
\label{dmetric}
ds^2 = \alpha^{2\over {\bar D}-2} \left \{ e^{2{\bar a}
{\bar\phi}}\left [ dt^2 - dy_{{\bar p}}^2 \right]
- e^{-{\bar\gamma}{\bar\phi}}dx_{{\bar n},\kappa}^2 \right \}
-\alpha^{-{({\bar D}-3)\over({\bar D}-2)}} e^{2{\bar\chi}} (dr^2 + dz^2 )
\ee
Then the equations of motion are:
\bea
\Delta \alpha &=& -2\Lambda \alpha B^2 + {\bar n}({\bar n}-1)\kappa
\alpha G\label{Dalpheq} \\
{\bar a}\left ( \Delta {\bar\phi} + \nabla {\bar\phi}
\cdot {\nabla\alpha\over\alpha} \right)
&=& -{{\bar n}({\bar n}-1)\over({\bar D}-2)}\kappa G \label{Dphieq} \\
\Delta {\bar\chi} + {\quarter} (\nabla {\bar\phi})^2 &=& - {\Lambda
B^2\over {\bar D}-2}
-{{\bar n}({\bar n}-1)\kappa G\over2({\bar D}-2)}\label{Dchieq} \\
{\partial_\pm^2\alpha\over\alpha} + {\half} (\partial_\pm{\bar\phi})^2
- 2 \partial_\pm {\bar\chi} {\partial_\pm\alpha\over\alpha}
&=& 0 \label{Dconsteq}
\eea
It is important now to emphasize that although (\ref{metric}) and 
(\ref{dmetric}) are different,
the constraint equations (\ref{consteq}) and (\ref{Dconsteq})
are unchanged, as is the form of the
$\alpha$ equation. Furthermore the expressions for
$B^2$ and $G$ are now
\bea
B^2 &=& e^{2\bar \chi} \alpha^{-({\bar D}-3)/({\bar D}-2)} \label{DBform}\\
G &=& B^2/C^2 = e^{2\bar \chi} e^{\gamma {\bar\phi}}
\alpha^{-({\bar D}-1)/({\bar D}-2)}\label{DGform}
\eea

With the new definition of the variables
(\ref{Aphidual}-\ref{Bchidual}) there is therefore a natural parallel,
or duality, between the
$\kappa = 0$, $\Lambda \neq 0$ system and
the $\Lambda = 0$, $\kappa \neq0$ barred system. Indeed by performing
the mapping,
$$
\chi\leftrightarrow {\bar \chi}, \qquad \phi\leftrightarrow {\bar\phi}
$$
and formally identifying,
\be
\label{duality}
\qquad n\leftrightarrow -{\bar D}+2,
\qquad \kappa\leftrightarrow -{2\Lambda\over ({\bar D}-1)({\bar D}-2)}
\ee
the expressions for $B^2$ and $G$ are exchanged from
(\ref{Bform})-(\ref{Gform}) to (\ref{DBform})-(\ref{DGform}) and vice-versa.
Therefore the field equations (\ref{alpheq}-\ref{consteq})
and (\ref{Dalpheq}-\ref{Dconsteq}) are exactly equivalent. In practical
terms this means
that once we have a set of solutions (in arbitrary dimension $D$) for the
$\kappa = 0$, $\Lambda \neq 0$ ${\bar\phi}$-Weyl system we can, using
(\ref{duality}), obtain the set of solutions for the
$\Lambda = 0$, $\kappa \neq0$ $\phi$- Weyl system and vice-versa. The
duality relates vacuum $D$-dimensional solutions with an
internal curvature source
$\kappa$, to constant curvature $\Lambda$ spacetimes of dimension
${\bar D}$. Of course this is not a relation between
physical spacetimes, since for $n\geq2$, ${\bar D}\leq0$. However,
many of the equations that follow, and their corresponding solutions,
are written formally in terms of $n$ and $D$ and so can be
dualized. The same of
course holds for the $n+2$-dimensional scalar field counterparts of
these Weyl spacetimes. The
duality in that case relates free scalar field solutions with internal
curvature $\kappa$ to scalar field solutions with a Liouville
potential. We shall be making use of this in section \ref{sec:wlam} 
(for an explicit example see section \ref{ssec:wlamc2}).

\section{Vacuum spacetimes with a flat transverse space}

If $\kappa$ and $\Lambda$ both vanish, this is clearly of the
Emparan-Reall \cite{er}
form, and writing $\phi = 2\lambda$, we see that the equations are
identical to the four-dimensional Weyl-vacuum equations. Thus any
four-dimensional vacuum solution automatically generates a
higher dimensional solution. While these solutions are
implicitly contained in \cite{er}, it is useful to consider
a few simple examples to understand the appropriate generalization
of the four-dimensional Weyl solutions.

First of all, note that the metric (\ref{genwmet}) as written is Poincare
invariant in the $(t,y)$ directions. This means that any solution has
the interpretation of a boost symmetric or {\it cosmic} $p$-brane
(in the sense of \cite{CPB}). So let us consider the case of $n=1$,
where the space orthogonal to the $p$-brane is three dimensional.
Since $n=1$, naturally $\kappa=0$, and so we can write
\be
\alpha = r \;\;,\;\;\;\;
\phi = 2\lambda_4 \;\;,\;\;\;\;
\chi = \nu_4
\ee
or
\be
\label{pext4}
ds^2 = \left( e^{2\lambda_4} \right )^{\sqrt{2\over(p+1)(p+2)}} [dt^2 - dy_p^2]
- \left ( e^{-2\lambda_4} \right ) ^{\sqrt{2(p+1)\over(p+2)}} \left [
e^{2\nu_4} (dr^2 + dz^2) + r^2 d\theta^2 \right]
\ee

A natural four-dimensional potential to consider is that of the
Schwarzschild black hole, (\ref{Schlam}), which, after unravelling back
to a spherical coordinate system via (\ref{w2sch}), gives
\be
ds^2 = \left ( 1 - {2M\over\rho} \right )^{\sqrt{2} \over \sqrt{(p+1)(p+2)}}
[dt^2 - dy_p^2] - \left ( 1 - {2M\over\rho} \right )^{-\sqrt{2(p+1)\over(p+2)}}
[d\rho^2 + \rho(\rho - 2M) d\Omega_{I\!I}^2]
\ee
in agreement with the results (see eqn (3.8)) in \cite{CPB}. This method
therefore provides a separate confirmation that the appropriate
Poincar\'e symmetric $p$-brane solutions are of this singular form.

The existence of a singular solution seems disturbing after our experience
of four-dimensional gravity, where singularities are mostly cloaked by
event horizons. However, in higher dimensions with extended solutions,
the natural nonsingular black branes of Horowitz and Strominger are
unfortunately generally unstable \cite{GL}, and it seems likely that
the true stable extended solution simply is this singular one. Indeed,
the conical deficit of the extended string solution in four dimensions is
strictly speaking singular, in that the Ricci scalar has a delta-function
singularity at the string core. This is nonetheless tolerated as field theory
is well defined on the remaining spacetime -- wave operators are self-adjoint
and the propagator is well defined. In this case, these higher dimensional
extended solutions, while containing null singularities, also have the
property that wave operators are self-adjoint, with Green's functions
being well defined \cite{CER}.
It seems likely therefore that despite our qualms,
these singular solutions should be accepted as legitimate additions to
the family of extended gravitating solutions and indeed perhaps
the only appropriate ones. It is in fact quite likely that
these are the endpoints of any black string instability.

We can also construct more complicated black brane metrics, such as
two $p$-branes held in equilibrium at a fixed distance apart by
$(p+1)$-brane conical deficits by simply inputting the Israel-Khan
potentials $\lambda_4$, $\nu_4$, from (\ref{iklam},\ref{iknu}).
However, perhaps a more interesting physical set-up to consider
is that of an accelerating $p$-brane obtained by putting in the
C-metric potential:
\bea
\label{cbrane}
ds^2 &=& \left ( {F(y)\over A^2 (x+y)^2}
\right)^{\sqrt{2}\over\sqrt{(p+1)(p+2)}}
\left[ dt^2 - dy_p^2 \right] \nonumber \\
&-& \left ( {F(y)\over A^2 (x+y)^2} \right)^{1 - \sqrt{2(p+1)\over(p+2)}}
{1\over A^2(x+y)^2} \left [ {dy^2\over F(y)} + {dx^2\over G(x)}
+ G(x) d\phi^2\right]
\eea
The interesting point to note about this metric is
that it is now not only singular at the $p$-brane horizon,
but also at the acceleration horizon. In part, this is
because of geometry -- Rindler spacetime is a transformation of flat
space, and it is not possible to slice flat space in such a way
as to have a flat accelerating brane. Another way of putting
this is to see that using the Rindler potentials (\ref{canrnw})
for (\ref{pext4}) gives a non-flat metric. In order to have
zero Weyl curvature, we must have curvature on the $p$-brane, \ie\
an Einstein-de Sitter metric. 

Does this mean that (\ref{cbrane})
has no physical significance? This is an interesting question.
In the case of domain walls or global vortices, imposing a Poincar\'e
symmetry on the defect leads to a singular metric, whereas allowing a
dS worldbrane removes this singularity. This is also the case for
a pure gravitating ``spherically symmetric'' $p$-brane \cite{rinfl}.
Although we are interested here in general axisymmetric pure
gravity solutions, the fact that formerly regular horizons become
singular as dimensionality is altered, and also the possibility of
removing those singularities by altering the intrinsic geometry
of the brane is a key point, and one we will return to later. It
may well be that if we wish a cosmic censor to be operative, we
are severely restricted as to the types of geometry we can consider.
Unfortunately, only the existence of the regular inflating
branes of \cite{rinfl} are known, not an explicit solution.

We can also extend the C-metric in arbitrary dimensions for $\kappa=\Lambda=0$. To do this, rather
than bringing the C-metric to the Weyl form, we coordinate transform
(\ref{dmetric}) for $D=4$ and $\Lambda=0$,
\be
\label{dmetrictr}
ds^2 = \alpha \left \{ e^{\bar\phi}dt^2
- e^{-{\bar\phi}}dx^2 \right \}
-\alpha^{-1/2} e^{2\bar \chi} ({df^2\over f'^2} + {dg^2\over g'^2} )
\ee
in a ``canonical form'' where we set
\be
\left({df\over dr}\right)^2=f'^2=F(f),\qquad \left({dg\over dr}\right)^2
=g'^2=G(g)
\ee
with $F$ and $G$ given by (\ref{cmetric1}).
Then identifying (\ref{cmetric}) and (\ref{dmetrictr}) we get,
\be
\label{an}
\alpha={f'g'\over A^2(f+g)^2},\qquad {\bar\phi}=\ln{f'\over g'}
\ee
and the field equations (\ref{Dalpheq}-\ref{Dconsteq}) are satisfied
with,
\be
e^{2\chi}={\sqrt{f'g'}\over A^3(f+g)^3}
\ee
Of course for the moment we have achieved nothing new,
this is just the C-metric for $D=4$ 
in the variables we defined in the previous section.
However in these variables the solution can now be extended to arbitrary
dimension $D$ {\it as long as} $\Lambda=0$ and $\kappa=0$.
The solution takes the rather complicated form,
\bea
\label{dentist}
ds^2&=&\left({1\over A(f+g)}\right)^{4\over D-2}
\Bigl\{ -{(FG)^{-{D-4\over 4(D-2)}}\over [A(f+g)]^{D-4\over D-2}}
\left({df^2\over F}+{dg^2\over G} \right)
\nonumber\\
&+&(FG)^{1\over D-2}
\left[ ({F\over G})^a [ dt^2 - dy_p^2]-({F\over G})^{\gamma\over 2}
dx_n^2\right] \Bigr\}
\eea
This solution is regular only for $D=4$. Indeed note then  
from (\ref{alpheq}-\ref{consteq}) that we can
have a source term $\kappa=1$ (and thus a well defined horizon) 
since $n=1$. For $D>4$ the requirements of planar
topology, and the absence of a cosmological constant, do not
permit the screening of a singularity, analogous to the case of
the ordinary planar symmetric four-dimensional spacetime
\cite{taub} which is also singular.

\section{Weyl metrics with subspaces of constant curvature}

In the general vacuum case, where $\Lambda = 0$, but $\kappa=1$, we
would ideally like to classify all solutions. Note that this case is
closely related to braneworlds with bulk scalars via the dimensional
reduction discussed earlier. Clearly, there is a large class of solutions
which are effectively one-dimensional following from earlier work in
braneworld scalar solutions \cite{BWS}, 
as well as a more general (though
still essentially one-dimensional) analysis in \cite{christos}.
These cases can be considered as special within the context of a more general
analysis which we now present.

The key is to  consider the $\phi$ equation (\ref{phieq})
which is now homogeneous. This equation states that
$*\alpha {\bf d} \phi$ is a closed form. This in turn suggests
that writing $\phi=\phi(z)$
should pick up a large class of solutions. Inputting this form for $\phi$
into (\ref{phieq}) implies that $\alpha$ is separable:
\be
\label{sep1}
\alpha = f(r)g(z),
\ee
where
\be
\label{sep2}
g(z) = c/{\dot\phi}
\ee
and $c$ is an arbitrary nonzero constant for ${\dot\phi}\neq0$, if
${\dot\phi}=0$ we set $c=0$, $g\equiv1$. Therefore under the hypothesis
$\phi=\phi(z)$
we can find the general solution to the field equations once we solve
for $f$, $g$ and $\chi$
using (\ref{alpheq}-\ref{consteq}).
We have three classes of possible solutions: Class I with $f'=0$, Class II
with $g'=0$, or, Class III where neither $f'$ or $g'$ vanish. Let us deal
with these in turn.

\subsection{Class I solutions}

Given that $f'=0$,
Class I solutions are characterised by the fact that they
depend only on one variable. Equivalently we can note that
Class I solutions manifestly have an extra Killing vector, 
therefore, we can expect solutions of greater than Weyl
symmetry appearing in this class.  For ${\dot\phi}\neq 0$ 
(\ref{sep2}), after some algebra we obtain a static
$p$-brane solution:
\be
ds^2 = V(\xi)^{-{n^2\over M (n-1)(p+1)}} (dt^2-dy^2_{p})
-V(\xi)^{{n\over M (n-1)}-{\mu \over M}}dr^2-V(\xi)^{{\mu+n+M
\over (n-1) M}}\left(
{d\xi^2 \over V(\xi)}+\xi^2 dx_{n,\kappa}^2\right)
\label{class11}
\ee
of axial symmetry with  $\xi$ the radial coordinate. Here of course we are
restricted to $n\neq 1$.
We have defined a spacetime dimension dependent parameter
\be
\label{mass}
M^2=\mu^2+{n^2(n+p+1)\over (n-1)(p+1)},
\ee
and $\mu$ is an integration
constant.
The potential $V(\xi)$ reads,
\be
\label{pot}
V(\xi)=1+{M \over \xi^{n-1}}
\ee
where $M$ is not necessarily positive. Note that while 
class I solutions are  asymptotically flat, the solution has
curvature singularities at $\xi=0$ and $V(\xi)=0$. This is the
most general axisymmetric solution depending on one coordinate
only, and generalizes the cosmic $p$-brane solutions of 
\cite{CPB}{\footnote{It is worth noting here that $\xi=0$ is an event
horizon for $c\neq 0$ only if $\mu={n\over n-1}$ and $p=0$.
Then the solution reduces to the
black string solution given by (\ref{unstable})}}.

Analytic continuation between the component independent
coordinates can yield
different solutions. With little effort for example
we can obtain a de-Sitter or inflating
$(n-1)-$brane solution,
\be
ds^2 = V^{{\mu+n+M\over (n-1) M}}\left[\xi^2
(d\tau^2-e^{2\sqrt{\kappa}\tau}dx^2_{n-1})
-{d\xi^2 \over V}\right]\nonumber -V^{{n\over M (n-1)}-{\mu\over M}}dz^2-
V^{-{n^2\over M (n-1)(p+1)}} dy^2_{p+1}
\label{class12}
\ee
where the curvature radius of the spherical sections (\ref{class11})
translates into the de-Sitter expansion of the $(n-1)-$brane.

Alternatively, to obtain the class I  Thorne vacuum solution simply
take $z\leftrightarrow i\tau$, $t\leftrightarrow iy$,
\be
ds^2 = V^{{n\over M (n-1)}-{\mu\over M}}\kappa d\tau^2-
V^{-{n^2\over M (n-1)(p+1)}} dy^2_{p+1}-V^{{\mu+n+M \over (n-1) M}}\left(
{\kappa d\xi^2 \over V}+\xi^2 dx_{n,\kappa}^2\right)
\label{class1}
\ee
It is now apparent that these solutions are the extensions of the cosmic
$(p+1)-$brane solutions \cite{CPB} (see also \cite{CER} for
an application of these spacetimes to braneworld models).
Indeed here the cosmic $(p+1)-$brane exhibits `cosmological'
rather than Poincar\'e symmetry. To
recover the flat case \cite{CPB}
one simply equates the metric components obtaining
\be
\mu_{flat}={n(n+p+1)\over (n-1)(p+1)}
\ee
The solutions (\ref{class1}) are asymptotically flat and singular at the
origin just like their flat counterparts \cite{CPB}.

Thorne class I solutions are all related
via (\ref{Dact}) to a free
scalar field spacetime of dimension $d=n+2$.
The class I solution (for arbitrary $c$) reads,
\be
\label{dilaton1}
ds^2=V^{-\mu\over M}\kappa dt^2-V^{{\mu -c(n-2)M \over (n-1) M}}
\kappa d\xi^2
-V^{\mu+cM \over M(n-1)} \xi^2 dx_{n,\kappa}^2
\ee
with,
\be
\phi={ c n\over M(n-1)\gamma}\, \ln |V|
\ee
For $\phi=$const.\ the class I solution (\ref{class1}) is 
simply the $d$-dimensional
Schwarzschild black hole extended in $(p+1)$ dimensions
\be
\label{unstable}
ds^2=\left(\kappa-{\mu\over \xi^{n-1}}\right)dt^2
-{d\xi^2\over\left(\kappa-{\mu\over \xi^{n-1}}\right)}-\xi^2 dx_{n,\kappa}^2
-dy^2_{p+1}
\ee
with ADM mass proportional to $\mu$ \ie\ the Horowitz-Strominger
black $(p+1)-$brane.  These solutions are the only regular
solutions for this class of metrics, however,
it has been shown that they are unstable \cite{GL} and hence unphysical.

\subsection{Class II solutions}

For Class II solutions, \ie\ $g'=0$, the $\phi$ field depends linearly
on $z$, but every other variable  depends only
on $r$. These solutions are therefore minimally two-dimensional.
The field equations then reduce to a single third order
non-linear differential equation for $f(r) = \alpha / g$:
\be
\label{completely}
{f''\over f} - {f'\over f} \left ( {f'\over nf} + {f'''\over f''} \right )
= {1\over 2g^2}
\ee
Without loss of generality we can set $g=1$, and by writing
$X=f'/f$, $Y = f'''/f''$, this equation can be recast as a two
dimensional dynamical system which allows us to analyse the 
general form of the solution:
\bea
X' &=& XY - {(n-1)\over n} X^2 + {1\over2} \label{2dwdsx}\\
Y' &=& {(n-2)\over n^2}X^2 - {2\over n} XY - {(n+2)\over 2n}
\label{2dwdsy}
\eea
This has no finite critical points, however, there are four very clear
asymptotes, two for increasing $r$ and two for decreasing $r$. The
phase plane is plotted in figure \ref{fig:weyl} for $n=2$, although
the picture is qualitatively the same for all $n$.
\FIGURE{
\includegraphics[width=14cm]{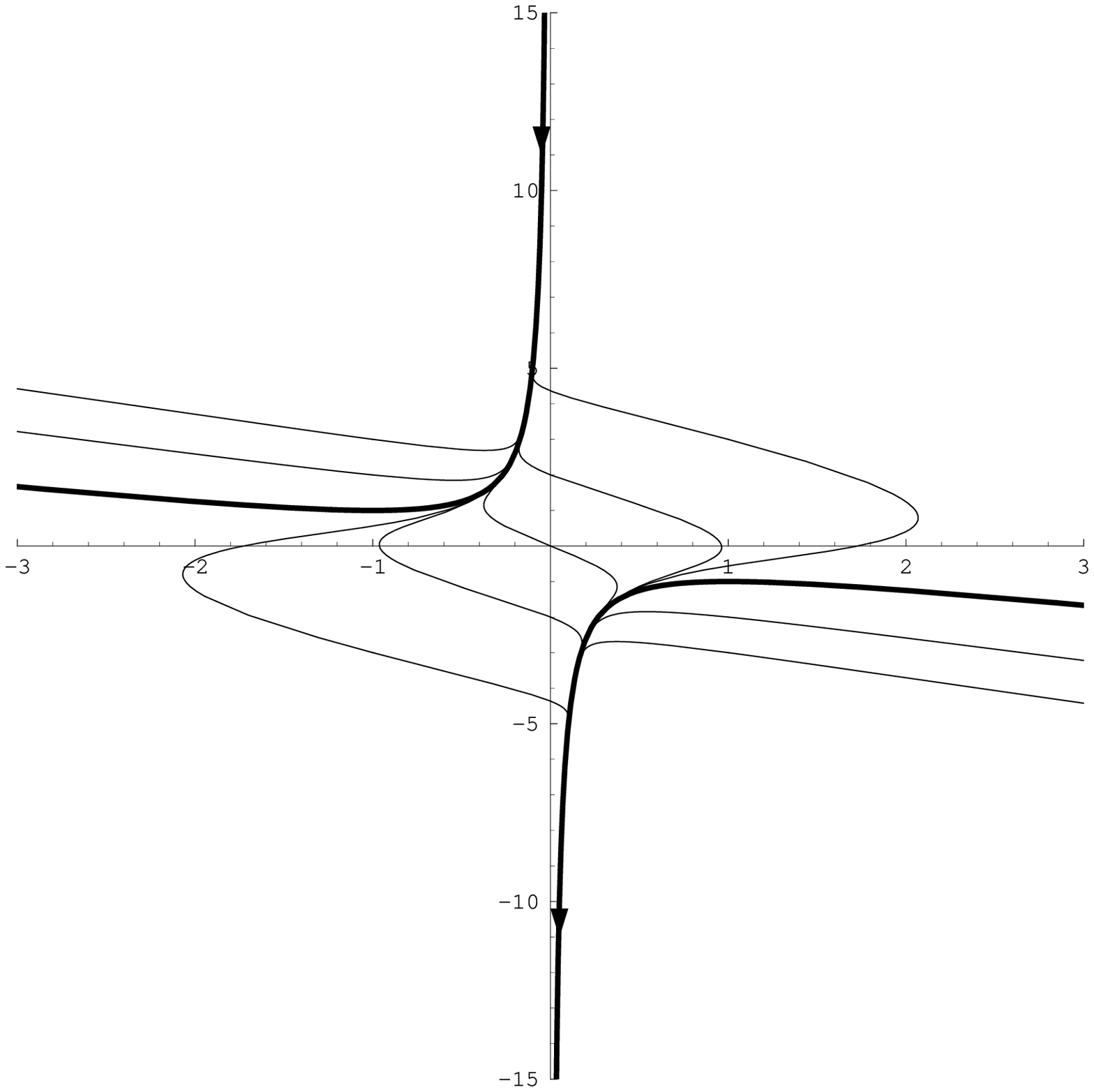}
\caption{The Weyl class II phase plane. The thick lines represent
the invariant hyperboloid ${\cal C} = XY + {X^2\over2} + {1\over2}=0$
which corresponds to $\kappa=0$. The positive $\kappa$ region lies
inbetween the two branches of ${\cal C}$.}
\label{fig:weyl}
}

Note that since $\alpha = f$, we must have $f'' = f (XY+
{X^2\over n} + {1\over2}\neq0$ for a solution of the Einstein
equations with $\kappa\neq0$. The phase plane therefore splits into
two regions corresponding to positive and negative $f''$. These
are separated by the invariant hyperboloid $Y = -X/n-1/2X$, shown
as the thick line in figure \ref{fig:weyl}. The connected region
between the two branches of this hyperboloid corresponds to positive
$\kappa$.

To find these asymptotes, note first that by plotting the
isoclines we see that $Y \sim {-1\over2X}$ is an
asymptotic solution which, solving for $X$, hence $f$ and $\alpha$,
corresponds to the metric
\be
\label{isoasy}
ds^2 = e^{2az} \left [ dt^2 - d{\bf y}_p^2 \right]
- e^{-2(p+1)az/n} \left [ |r|^{2\over n} d\Omega_n^2
+ {e^{-r^2/4}\over n(n-1) |r|^{n-2\over n}} (dr^2 + dz^2) \right ]
\ee
for large $|r|$. In other words, this asymptotic solution
corresponds to a singular infinity for $|r|\to\infty$.

For the other asymptotes, note that for large $X$ and $Y$, a solution
to (\ref{2dwdsx},\ref{2dwdsy}) must have $Y = \lambda_\pm X$,
where $\lambda_\pm
= [(n-3)\pm(n-1)]/(2n)$. The first of these, $\lambda_+$ corresponds
to a separatrix between solutions which asymptote (\ref{isoasy})
and those which asymptote $Y=\lambda_-X = -X/n$, which correspond
to the metric
\be
\label{assol}
ds^2 = e^{2ac_0z} \left [ dt^2 - d{\bf y}_p^2 \right]
- e^{-2(p+1)ac_0z/n} \left [ R^2 d\Omega_n^2
+ {dR^2\over 1 + {\mu\over R^{n-1}}} + \left ( 1 + {\mu\over R^{n-1}}
\right ) dz^2 \right ]
\ee
for $R\to0$, where $\mu>0$ and $c_0$ are integration constants.
Although this solution is reminiscent of the Euclidean black hole
cigar, since $\mu>0$ and $z$ is not periodically identified this
solution is singular as $R\to0$.

Now that we have these asymptotic forms, we can see that
there are two distinct types of class II spacetime solutions,
one for $\kappa=1$, which asymptotes (\ref{isoasy}) for
both large negative and large positive $r$, and
one for $\kappa = -1$, which
for small finite $r$ looks like the Euclidean Schwarzschild
style solution (\ref{assol}) and asymptotes (\ref{isoasy}) for large $r$.
(The trajectory starting from (\ref{isoasy}) and terminating on
(\ref{assol}) is simply this spacetime reversing $r$.)

If however, we are dealing with a Thorne rather than a Weyl metric,
then the sign of the RHS of (\ref{completely}) changes, and hence 
the signs of the constant terms in (\ref{2dwdsx}),(\ref{2dwdsy}) change. 
The main effect of this is that it introduces a pair of critical
points $P_\pm = \pm (\sqrt{n\over2}, \sqrt{n\over2})$, $P_+$ 
an attractor, $P_-$ a repellor, which are focal in nature for $n<8$.
The phase plane is shown in figure \ref{fig:thrn} for $n=2$.
\FIGURE{
\includegraphics[width=14cm]{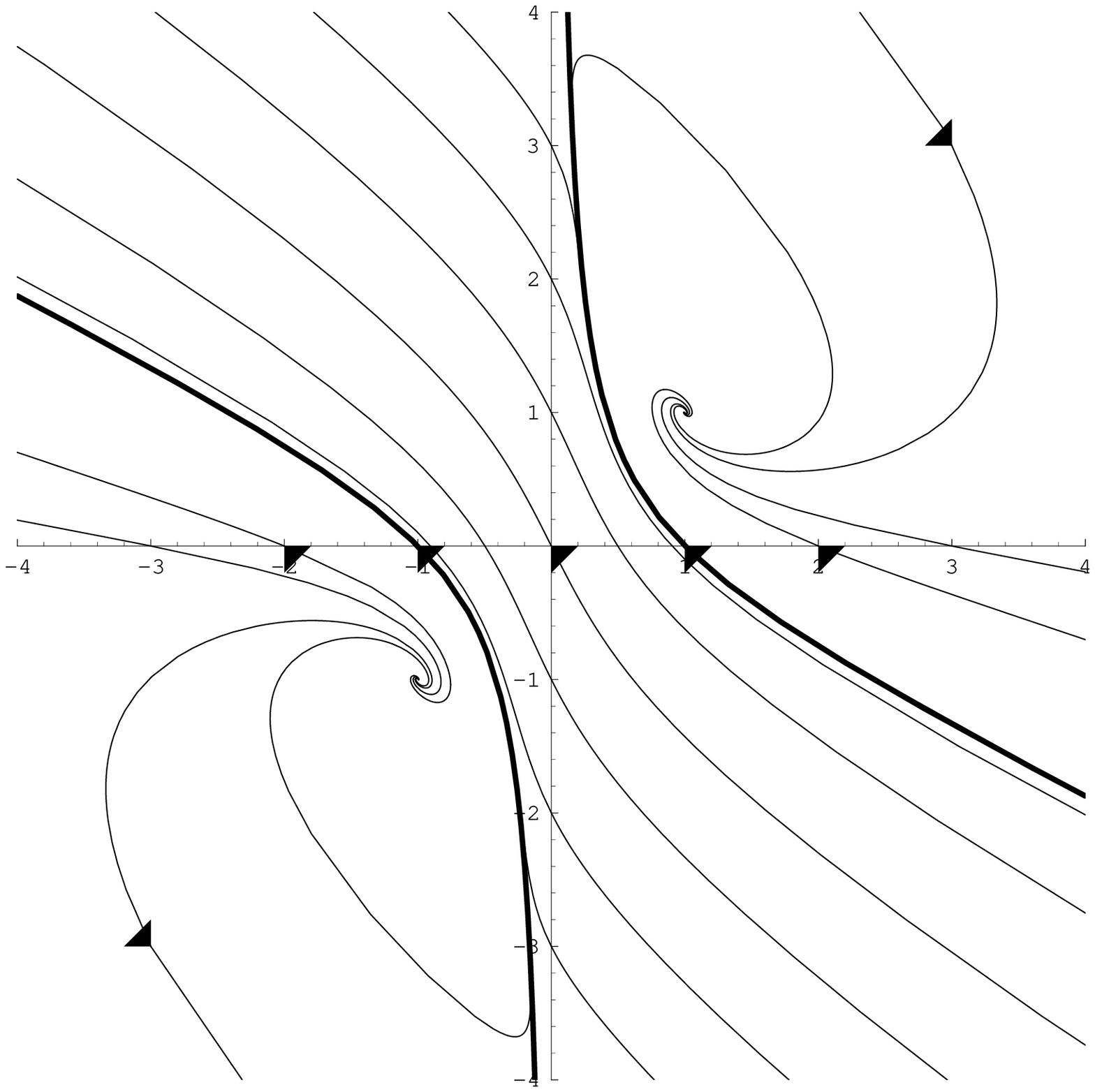}
\caption{The Thorne class II phase plane for $n=2$. Once again the thick
lines are the invariant hyperboloid ${\cal C} = XY +{X^2\over2} 
-{1\over2}=0$, with the region inbetween the two branches now 
corresponding to negative $\kappa$. The critical points have 
$\kappa=1$, and the attractor solution $P_+$ therefore corresponds to the
generic asymptotic spacetime.}
\label{fig:thrn}
}
The critical point solution is obtained by setting 
$f=e^{r}$ (restoring the constant $g$ in (\ref{completely})),
\be
\label{2dthorne}
ds^2=e^{2r\over n}e^{2t\sqrt{p+1}\over n\sqrt{n+p+1}}\left({dt^2-dr^2\over
n(n-1)\kappa} -dx_{n,\kappa}^2\right)
-e^{-2t\over \sqrt{(n+p+1)(p+1)}}dy^2_{p+1}
\ee
The coordinates $t$ and $r$ vary on the whole real line and the
 metric is singular as $t \rightarrow -\infty$ and
as $r \rightarrow -\infty$.
For the dilaton spacetime we have simply,
\be
ds^2=e^{2r\over n}\left({dt^2-dr^2\over n(n-1)\kappa}
-dx_{n,\kappa}^2\right)
\ee
\be
\phi=-\sqrt{2\over n}t+\phi_0
\ee
the linear dilaton metric solution.
The general form of the Class II Thorne metric is given by,
\be
\label{2dthornegen}
ds^2={e^{2t\sqrt{p+1}\over n\sqrt{p+n+1}}\over
n(n-1)\kappa}f''(r)
(dt^2-dr^2)
-f(r) e^{2t\sqrt{p+1}\over \sqrt{p+n+1}}d\Omega_{n}^2
-e^{-2t\over \sqrt{(p+n+1)(p+1)}}dx^2_{p+1}
\ee
where $f$ is a solution of (\ref{completely}).

\subsection{Class III solutions}

In the case of class III, where both $f$ and $g$ are nontrivial,
the field equations (\ref{alpheq}-\ref{consteq})
give after some algebra two possible families of equations
for $f$ and $g$:
\bea
f^{\prime^2} = f_0 f^{2-2/n} - c_0 f^2 &~& g^{\prime^2} =
{n\over2(n+1)} + c_0 g^2 \label{fam1} \\
{\rm or} \;\;\;\;
f^{\prime^2} = d_0 f^2 &~& g^{\prime^2} = {n\over2(n-1)} + g_0 g^{2-2/n}
- d_0 g^2 \label{fam2}
\eea
The constants $f_0$ and $g_0$ are both different from zero since
otherwise $\Delta\alpha =0$.

The first family of equations (\ref{fam1}) are readily solved (with
$c_0\neq 0$) to give
\be
\label{fam11}
f = \left ( {f_0\over c_0} \right)^{n\over2} \sin^n {\sqrt{c_0}r\over n}
\qquad g = {\sqrt{n}\over\sqrt{2c_0(n+1)}} \sinh \sqrt{c_0}z
\ee
or, writing
\be
\xi^n = {f_0^{n\over2} \sqrt{n}\over c_0^{(n+1)\over2} \sqrt{2(n+1)}}
\left[1+\cosh\sqrt{c_0}z\right] = {M\over 2} \left[1+\cosh\sqrt{c_0}z\right]
\ee
gives the metric
\bea
\label{massivepbrane}
ds^2 &=& \left ( 1 - {M\over\xi^n} \right
)^{\sqrt{(n+1)\over(p+1)(p+n+1)}}
\left [ dt^2 - dy_p^2 \right] \nonumber \\
&-&  \left ( 1 - {M\over\xi^n} \right )^{\left(1-\sqrt{(n+1)(p+1)\over
(p+n+1)}\right)/n}
\left[ \xi^2 d\Omega^2_{n+1} + {d\xi^2\over 1-{M\over\xi^n}} \right]
\eea
which is in fact a massive boost symmetric $p$-brane solution \cite{CPB}.
The case $p=0$ corresponds to the Schwarzschild solution in $n+3$
dimensions. Notice here, that unlike the previous cases, $n=1$ is permitted.
In the case that $c_0<0$, one gets a hyperbolic black brane solution in
the region \emph{inside} the event horizon. Furthermore using (\ref{metric})
and by analytic continuation
we obtain the ``dimensionally reduced'' free scalar field metric,
\be
ds^2 =\left( 1 - {M\over\xi^n} \right )^{1/n}
\left[ \xi^2 (d\tau^2- e^{2\sqrt{\kappa}\tau} dx_n^2) -{d\xi^2\over 1-{M\over\xi^n}}
 \right]
\label{worm}
\ee
\be
\phi=\sqrt{n+1\over 2 n} \ln\left( 1 - {M\over\xi^n} \right )+\phi_{0}
\ee
which is singular at $\xi=0$ and $\xi^n=M$. 
Note that this metric is conformally an angular analytic continuation
of a Schwarzschild wormhole.
The presence of the scalar field here however,
renders the usual horizon $\xi^n=M$ singular.  

For $c_0=0$, (\ref{fam1}) is easily integrated and
we get a different type of solution:
\bea
\label{rindler}
ds^2 &=& \left ( z \right ) ^{2\sqrt{n+1}\over\sqrt{(p+1)(p+n+1)}}
\left[ dt^2 - dy_p^2 \right]\nonumber \\
&-& \left ( z \right ) ^{{2\over n}\left(1-\sqrt{(n+1)(p+1)\over
(p+n+1)}\right)}
\left [ dz^2 + dr^2 + r^2 d\Omega_n^2 \right]
\eea
This is reminiscent of a Rindler metric, although if $p\neq0$ it is not a
flat solution and in fact has singularities at $z=0$ and $z=\infty$.
For $p=0$ writing $T = z\sinh t$,
$Z = z\cosh t$ we get the Minkowski metric. This shows that this is
a flat space solution, giving the Rindler metric in $n+3$ dimensions tailored
to an accelerating particle.

The second family of solutions do not have an interpretation
in terms of known spacetimes. For simplicity let us examine (\ref{fam2})
for $n=2$, for which $g$ is readily integrated.
Writing $R = \sqrt{g_0/d_0} e^{\pm \sqrt{d_0}\ r/2}$,
$2\theta = \sqrt{d_0} z - \pi/2$, and $\lambda_0 = 2\sqrt{d_0}/g_0$, we have:
\be
e^\phi = G(\theta) = {(1+\lambda_0+\sqrt{1+\lambda_0^2})\sin\theta
+ (1-\lambda_0-\sqrt{1+\lambda_0^2})\cos\theta \over
(1-\lambda_0+\sqrt{1+\lambda_0^2})\sin\theta
+ (1+\lambda_0-\sqrt{1+\lambda_0^2})\cos\theta}
\ee
with the metric
\bea
ds^2 =  G^{2a} \left [ dt^2 - d{\bf y}^2 \right ]
&-& G^{-a(p+1)} \Biggl \{ dR^2 \\
&+& R^2 \left [ d\theta^2 + \left ( \sqrt{1+\lambda_0^2} \sin^2\theta 
+ {1-\sqrt{1+\lambda_0^2}\over2} \right)
d\Omega_{I\!I}^2 \right ] \Biggr \}\nonumber
\eea
$\lambda_0\to0$ is clearly the flat space limit. 
For nonzero $\lambda_0$ however,
we have an angular distortion of the spacetime, with singularities at
$\sin^2\theta = {\sqrt{1+\lambda_0^2}-1\over 2\sqrt{1+\lambda_0^2}}$,
one of which, with $G=0$ is a null singularity, and the other an
asymptotic singularity with $G\to \infty$.

\section{Weyl metrics in a constant curvature spacetime}\label{sec:wlam}

Let us now consider the case where both subspaces are
of planar topology ($\kappa=0$) and,
in contrast, switch on  a
cosmological constant $\Lambda$. Using the dual field
equations (\ref{Dalpheq}-\ref{Dconsteq}) we can obtain as before three
classes of solutions. As before, if we write $\phi=\phi(z)$, then 
for $\kappa=0$ the general solution can be found in the form
$\alpha=f(r)g(z)$, with $g$ defined by (\ref{sep2}). 
As before for class I we have $f'=0$, for class II $g'=0$,
and for class III $f'\neq 0$ and $g'\neq 0$. In the subsections that
follow we use extensively the duality relation (\ref{duality}) to map
the $\kappa\neq 0$ solutions to the
cosmological constant solutions, $\Lambda\neq 0$.
Here, unlike the case ($\kappa\neq 0$)
studied in the previous section, we can set $D=4$ finding the 
four-dimensional
versions of the solutions with a cosmological constant.

\subsection{Class I solutions}

The class I solution for $c\neq 0$ reads,
\be
\label{weylclass1}
ds^2=\xi^2 V^{\mu+M\over (n+p+2)M}\left[
V^{-n(n+p+1)\over M(p+1)(n+p+2)}(dt^2-dy^2_{p})-V^{-\mu\over M}dz^2-
V^{(n+p+1)\over (n+p+2)M}dx_n^2\right] -{d\xi^2\over k^2\xi^2 V}
\ee
with
\be
\label{mass1}
M^2=\mu^2+{n(n+p+1)^2\over (n+p+2)(p+1)},
\ee
The potential is now given by,
$$
V(\xi)=1+{M\over \xi^{D-1}}
$$
and we have set $-2\Lambda=(D-1)(D-2)k^2$. Solution (\ref{weylclass1})
asymptotically approaches adS spacetime and is singular for $V=0$, and
at $\xi=0$.
The Thorne spacetimes are  obtained with $z\leftrightarrow i\tau$,
$t\leftrightarrow iy$ where
for $\Lambda<0$ (adS)
the coordinate $\xi$ is timelike (and hence $t$ spacelike)
inbetween the critical points $\xi=0$ and $\xi=-(M)^{1\over D-1}$.
The coordinate is spacelike on the
exterior of the interval{\footnote{The situation is reversed for a De
Sitter cosmological constant}}. 

Starting from the Class I Thorne metric one can
make the connection with the dilaton
system in $d=n+2$ dimensions with potential
$V=2\Lambda e^{\gamma\phi}$ using (\ref{Dact}).
The solution corresponds to the Class I solutions found in \cite{christos}.

Starting from a Thorne metric and for
$c=0$ we obtain,
\be
\label{bh1}
ds^2=-(k^2\xi^2-{\mu\over \xi^{D-3}})dt^2+{d\xi^2\over k^2\xi^2-{\mu\over
\xi^{D-3}} }+\xi^2 dx^2_{D-2}
\ee
the planar topological black hole with cosmological constant. Note then
that this solution is dual to the black brane solution (\ref{unstable}).
The corresponding Weyl metric gives naturally
the Euclidean version of (\ref{bh1}). The dilatonic black hole solution
is then rather nicely seen to map to the black hole solution (\ref{bh1}) via
(\ref{Dact}), (see ref.\ \cite{mann}).

\subsection{Class II solutions}\label{ssec:wlamc2}

To obtain the Class II solutions we again make use of the
duality relation (\ref{duality}). In order to spell out the relation let
us consider to start with the simple case where $f=e^r$ is solution to
(\ref{completely}) treated in
section 5. We saw that in this case we can obtain the Class II vacuum solution
(\ref{2dthorne}) with components,
\be
\label{apply}
\alpha=e^r,\qquad \phi=-\sqrt{2\over n}t,
\qquad e^{2\chi}={e^{{n+1\over n}r}\over n(n-1)\kappa}
\ee
All we have to do now is set $n=-D+2$, $\kappa={-2\Lambda\over
(D-1)(D-2)}$
 and insert the above components in (\ref{dmetric}). The form of the scalar field in (\ref{apply}) dictates that we will have to analytically continue on applying the duality map (\ref{duality}). We then obtain with ease,
\be
\label{weyl2d}
ds^2= e^{2r\over {n+p+1}}\left(
e^{-2z\sqrt{n}\over (n+p+1)\sqrt{p+1}}(dt^2-dy^2_{p})
-e^{2\sqrt{p+1}z\over \sqrt{n}(n+p+1)} dx_{n}^2 \right)
-{dr^2+dz^2\over -2\Lambda} 
\ee
where note that now we have a Weyl solution in a
negatively curved spacetime. 
\FIGURE{
\includegraphics[width=14cm]{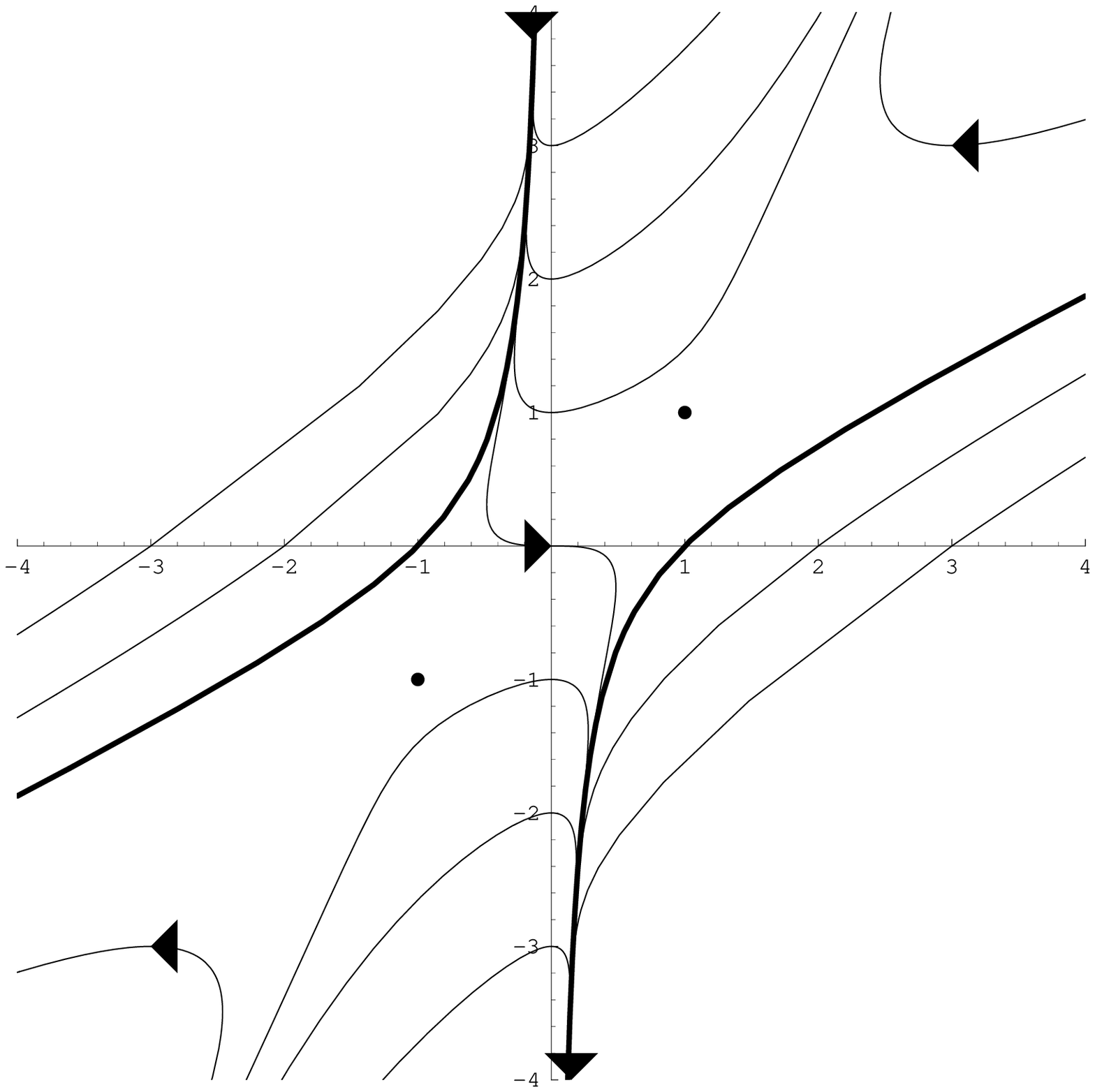}
\caption{The Weyl phase plane for class II solutions with a cosmological
constant. The thick lines are the invariant hyperboloid 
${\cal C} = XY - {X^2\over2} +{1\over2}=0$ which corresponds
formally to $\Lambda=0$. The connected region between the two
branches of ${\cal C}$ corresponds to negative cosmological
constant and hence the saddle critical point has $\Lambda <0$.}
\label{fig:lam}
}

Using again (\ref{duality}) we have that $f$ must satisfy the ODE,
\be
\label{Dcompletely}
{f''\over f} + {f'\over f} \left ( {f'\over (D-2)f}
- {f'''\over f''} \right )
= {1\over 2g^2}
\ee
As in the previous section we can set $g=1$, and by writing
$X=f'/f$, $Y = f'''/f''$, this equation can be recast as a two
dimensional dynamical system:
\bea
X' &=& XY - {(D-1)\over D-2} X^2 + {1\over2}\\
Y' &=& -{D \over (D-2)^2}X^2 + {2\over D-2} XY -
{D-4 \over 2(D-2)}
\eea

A representative phase plot is shown in figure \ref{fig:lam} for
$D=4$. Now it is the Weyl system which has two critical points
$P_\pm = \pm \left ( \sqrt{D-2\over2},\sqrt{D-2\over2}\right)$;
however, unlike the Thorne vacuum $\kappa=1$ spacetime, these critical
points are now saddles, and lie in the $\Lambda<0$ connected region
of the phase plane. The asymptotic solutions can be obtained from
(\ref{isoasy},\ref{assol}) by setting $n=-(D-2)$.

Generically the Class II solutions take the form,
\be
ds^2=-{f''\over f}{dr^2+dz^2\over
-2\Lambda} - f^{2\over D-2}\left(e^{2\sqrt{p+1}z\over
\sqrt{n}(n+p+1)} dx_{n}^2
-e^{-2z\sqrt{n}\over (n+p+1)\sqrt{p+1}}(-dt^2+dy^2_{p})\right)
\ee
where $f$ is a solution of (\ref{Dcompletely}).

\subsection{Class III solutions}

In the case of Class III where $f'\neq 0$ and $g'\neq 0$, the field
equations
(\ref{Dalpheq}-\ref{Dconsteq})
give after some algebra two possible sets of solutions
for $f$ and $g$:
\bea
f^{\prime^2} = f_0 f^{2{D-1\over D-2}} - c_0 f^2 &~&
g^{\prime^2} = {D-2\over 2(D-3)} + c_0 g^2  \\
{\rm or} \;\;\;\;
f^{\prime^2} = d_0 f^2 &~& g^{\prime^2} = {D-2\over 2(D-1)}
+ g_0 g^{2{D-1\over D-2}}
- d_0 g^2 \label{fam21}
\eea
with $\dot{\phi}(z)={1\over g(z)}$.

Taking the dual of solution (\ref{massivepbrane}), the first family
of solutions (\ref{fam1}) are easily obtained after a
bit of algebra for $c_0\neq 0$,
\bea
\label{class73}
ds^2 &=& {1\over (\sin kr)^2}\Bigl[
\xi^2 V^{{1\over {n+p+1}}({1+\sqrt{n(p+n)\over (p+1)}})} 
(dt^2-dy^2_p)\\
&-& \xi^2 V^{{1\over {n+p+1}}({1-\sqrt{(p+1)(p+n)\over n}})}
dx_n^2 - dr^2-{d\xi^2\over V k^2\xi^2}\Bigr]\nonumber
\eea
with the potential $V$ given by,
\be
V(\xi)=1-{M \over \xi^{n+p+1}}
\ee
This solution is singular for $\xi=0$ and $V=0$.
For $p=0$ however this solution is regular for $V=0$, and reads
\be
\label{blackadsstring}
ds^2 = (\cosh(kz))^2\left[
\xi^2 V
dt^2 -{d\xi^2\over V k^2\xi^2}- \xi^2 dx_n^2\right]-dz^2
\ee
where $z$ is now the proper distance coordinate. $V=0$ is now an 
horizon, and this metric 
describes an $(n+1)-$dimensional planar adS black hole embedded in an
$(n+2)-$dimensional adS spacetime. As $\xi\rightarrow 0$ (or $M=0$) 
we get an adS spacetime where the $n-$dimensional slicings are also of adS
geometry with the same curvature $k$. 
This solution interestingly is dual to the
usual Schwarszchild black hole (\ref{massivepbrane}). The black hole
singularity at $\xi=0$ is screened by what is now a horizon, $V=0$,
of planar topology.
Furthermore unlike the case studied in \cite{CHR} the solution is well
defined in the adS horizon since,
\be
(Riemann)^2 \sim k^4(1+\nu{M^2 (\sech(kz))^4\over \xi^6 }) 
\ee
where $\nu$ is some numerical coefficient
depending on the spacetime dimension.

For $c_0=0$ we obtain,
\bea
\label{class74}
ds^2 &=& {1\over r^2}\left[
z^{{2\over {n+p+1}}({1+\sqrt{n(p+n)\over (p+1)}})}
(dt^2-dy^2_p)- z^{{2\over {n+p+1}}({1-\sqrt{(p+1)(p+n)\over n}})}
dx_n^2-{(dr^2+dz^2)\over k^2}\right]
\eea
which for $p=0$ (\ref{class74}) reduces to adS spacetime. This solution
is quite naturally seen to be dual to the Rindler solution (\ref{rindler}).

The second family of solutions is given implicitly by,
\bea
\label{class73bs}
ds^2 &=& g^{2\over D-2}\Bigl[f^{2\over D-2}
\left(e^{2a\phi}(dt^2-dy^2_p)-e^{-\gamma\phi}dx_n^2
\right)\nonumber \\
&-& {{g_0\over (D-2)^2 k^2}}\left({df^2\over
d_0 f^2}+{dg^2\over {D-2\over 2(D-1)}
+ g_0 g^{2{D-1\over D-2}}
- d_0 g^2}\right)\Bigr]
\eea
where the component $\phi$ is given by,
\be
\phi=c\int {dg \over g\sqrt{{D-2\over 2(D-1)}
+ g_0 g^{2{D-1\over D-2}}-d_0g^2}}
\ee

\section{Linearization, the C-metric and braneworld black holes}

In this section, we would like to remark upon two open, related,
questions: where is the C-metric, and what can we say about black
holes on branes?
Before embarking upon this however, we would like to note that 
if a far-field description of a physical system is all that is
required, then there is an extremely straightforward linearization
prescription for the case where one of $\Lambda$ or $\kappa$ vanishes.
We will discuss $\Lambda=0$ for definiteness.

Start by observing that the flat space solution 
to (\ref{alpheq}-\ref{consteq}) is
\bea
\alpha &=& \alpha_0 =  r^n , \nonumber \\
\phi &=& 0 \\
\qquad e^{2\chi} &=& e^{2\chi_0} = r^{n-1} \nonumber
\eea
We now expand the metric functions in the usual way: $\alpha = \alpha_0
+ \varepsilon \alpha_1 + \ldots$. Since $\phi_0=0$, we 
see that the equations of motion actually decouple at first order:
\bea
\Delta \alpha &=& n(n-1) \alpha ^{-1/n} e^{2\chi}\nonumber \\
\Delta \chi &=& \frac{(n-1)}{2} \alpha^{-(n+1)/2} e^{2\chi} \\
\frac{\partial_\pm^2 \alpha}{\alpha} &=& 2 \partial_\pm \chi
\frac{\partial_\pm\alpha}{\alpha} \nonumber
\eea
and
\be
\Delta \phi + \nabla \phi \cdot \frac{\nabla\alpha}{\alpha} =0
\label{philap}
\ee
The first set of equations can actually be solved to all 
orders by a (Euclidean) black hole solution:
\be
\alpha = R(r)^n \qquad , \qquad e^{2\chi} = R^{n-1} - C
\ee
where $R(r)$ is defined implicitly by $r = \int n e^{-2\chi} d\alpha$.
Usually, the Euclidean black hole solution is characterised by the
periodic identification of the angular variable, in this case $z$.
However, this periodicity is imposed to make the solution regular at
the analytically continued event horizon, $R^n=C$. In our case, we
only require a far-field solution, therefore we do not need to
impose any periodicity on the $z$-coordinate, and simply note that
it is possible for $\alpha$ and $\chi$ to receive first order
corrections of this form, independent of the value of $\phi$.

The $\phi$-field is independently given to first order by
the solution of the Laplace equation (\ref{philap})
\be
\phi = -2a \int \frac{S({\bf r}')\ d^{n+2} {\bf r}'}{|{\bf r} -{\bf r}'|^n}
\ee
where $S({\bf r}')$ is a source term which can be thought of
as the matter source, and is exactly the higher dimensional
equivalent of (\ref{4lamsol}). 
 
As an example, consider the five-dimensional axisymmetric
metric with $p=0$ and $n=2$. Then for a line source lying between
$z_1$ and $z_2$, we have the corresponding linear $\phi$ potential:
\be
\phi = \frac{a}{r} \left [ \tan^{-1} \frac{z - z_2}{r}
- \tan^{-1} \frac{z-z_1}{r} \right ]
\ee
Not surprisingly this agrees to leading order with the black
hole solution, but any mass source ought to have its leading 
monopole order agreeing with the black hole.

In this case, to get the next order correction it is necessary to
return to a full system of PDE's, however, for some systems of
physical interest, the linear order term will be sufficient.

\subsection{The C-metric}

One of the original motivations of undertaking this study of static
axisymmetric metrics in higher dimensions was to try to generalise
the C-metric to greater than four dimensions. It would appear that
this attempt has not been successful. Let us first recap what the
C-metric physically represents. 

In four dimensions, the C-metric \cite{cmetric} represents two black
holes (in general charged) uniformly accelerating away from one 
another. In the usual ``relativists'' coordinates, (\ref{cmetric}), 
the metric depends on two variables, $x$ and $y$, roughly speaking
splitting into two pieces -- an angular part $(x,\phi)$, and
a `black hole' part $(t,y)$. 
As described in section \ref{subsec:Crin}, the $y$ coordinate, which pairs
up with the time coordinate, is effectively a radial variable in the
spacetime, and runs from the black hole to the acceleration horizon;
the $x$ coordinate is an angular variable, and runs 
from the conical singularity driving the acceleration to the direction
pointing directly towards the other black hole.

A more transparent form of the C-metric is obtained if we set
\be
{\bar t} = A^{-1}t \;\; , \;\; r = 1/Ay \;\; , \;\; {\rm and} \;\;\;
\theta = \int_x^{x_3} dx/\sqrt{G}
\ee
when
\be
ds^2 = [1 + Arx(\theta)]^{-2}
\left [ (1-{2m\over r} - A^2r^2) d{\bar t}^2 - {dr^2 \over
(1-{2m\over r} - A^2r^2)} - r^2 d\theta^2 - r^2 G d\phi^2
\right ].
\label{Kott}
\ee
This is almost conformally equivalent to the Kottler, or Schwarzschild
de-Sitter metric as might be expected from the acceleration
horizon and clearly shows the black hole nature of the spacetime,
as well as the existence of the acceleration horizon (which is the
de-Sitter horizon of the Kottler metric). From this form of the 
solution it is clear that we reduce to the Schwarzschild metric as
$A\to0$, whereas the canonical form (\ref{cmetric}) shows how we
reduce to the Rindler metric as $m\to0$.

The key characteristics of the C-metric we glean 
from this four-dimensional known solution are therefore the following:

\vskip 2mm

$\bullet$ We expect two horizons, one corresponding to the black hole,
and one corresponding to the acceleration horizon.

$\bullet$ We expect to have {\it two} parameters, representing
mass and acceleration, under which our solution reduces to the
known Rindler, or Schwarzschild, solutions as each parameter (or
two linearly independent combinations of said parameters) is set
to zero.

$\bullet$ We expect to have the higher dimensional equivalent of a
conical singularity meeting the black hole and event horizons for one
limit of the coordinates, which gives the physical impetus for
the acceleration of the black hole.

\vskip 2mm

Unfortunately, we did not find any solutions with two horizons,
and many of the horizons we did discover were in fact singular.
As we will discuss presently, we believe that singularities
may well be a necessary part of the higher dimensional C-metric.
In addition, a frustrating aspect of the work on the class III
solutions was that we found a family of solutions which contained
the Rindler and Schwarzschild metrics, but only those two metrics.
Although we have made various Ans\"atze which attempt to preserve
features common to both Rindler and Schwarzschild, all of these
attempts give only Rindler and Schwarzschild as possible solutions.

Dealing with the third bullet point we can make more progress however.
There is of course a ready
generalization of the conical singularity to higher dimensions --
the cosmic $p$-brane \cite{CPB} for $p=1$. These spacetimes are Poincar\'e 
boost symmetric like the conical singularity, and like the conical
singularity propagators are well defined on the spacetime \cite{CER}.
On the other hand, a striking feature of these spacetimes is that
they are singular -- the conical singularity turns into a null
curvature singularity. It is difficult to know whether this is a
feature that is to be accepted or avoided. If one wishes instead
a nonsingular boost symmetric source, one is forced to add intrinsic
curvature to the worldbrane -- to have a de Sitter type of induced
metric parallel to the source \cite{rinfl}. However, if one modifies
the geometry parallel to this source, then the Rindler limit would 
also contain such a modification, and the fact that Rindler spacetime
must be flat restricts us severely in the slicings that we are allowed
to take (see the discussion after (\ref{cbrane})).

An obvious first step in finding the C-metric would be to attempt to
find the generalization of the Aryal-Ford-Vilenkin (AFV) solution
\cite{AFV}. This solution in four dimensions represents a
cosmic string threading a black hole, and therefore a natural
extension would be to take a black hole in higher dimensions and
allow a cosmic $1$-brane to thread it. We have not found this
solution within our metrics, (most likely it will require
a numerical integration) but it would however give a first step of
intuition as to the likely geometry of the black hole horizon.

It has been suggested that allowing a singular $1$-brane to interact with
the black hole horizon would render the horizon singular: that the
null singularity of the $p$-brane would somehow `infect' the black 
hole horizon. One way of exploring this issue is to take the limit
in which the black hole has an infinitely large mass compared to the
$1$-brane. In other words, the $1$-brane intersects a planar Rindler
horizon\footnote{We would like to thank Roberto Emparan for
suggesting this limit.}. This is readily obtained from the $1$-brane
metric (\ref{massivepbrane}) by a Rindler transformation:
\be
t = Z \sinh T \qquad ; \qquad y = Z \cosh T
\ee
under which the metric becomes:
\bea
ds^2 &=& \left ( 1 - {M\over\xi^n} \right )^{\sqrt{(n+1)\over2(n+2)}}
\left [ Z^2 dT^2 - dZ^2 \right] \nonumber \\
&-&  \left ( 1 - {M\over\xi^n} \right )^{\left(1-\sqrt{2(n+1)\over
(n+2)}\right)/n}
\left[ \xi^2 d\Omega^2_{n+1} + {d\xi^2\over 1-{M\over\xi^n}} \right]
\eea
Although this is not in Weyl form, it is easy to see that the only 
singularity remains the null $1$-brane singularity. The Rindler
horizon $Z=0$ remains nonsingular.

Of course this is not a definitive result as to the nonsingularity
of the black hole event horizon. The Rindler horizon has
planar topology just as a very large mass black hole looks roughly
planar to a thin strong locally piercing it. A finite mass black 
hole horizon would of course have finite intrinsic curvature, and just
as the Poincar\'e $p$-brane null singularity becomes an horizon upon
adding intrinsic curvature, it is possible that here adding instrinsic
curvature to the Rindler horizon could yet render it a null
singularity.

\subsection{Black holes on branes}

Although without the C-metric we cannot give a completely satisfactory
resolution to the problem of braneworld black holes, we would like to
draw attention to a pragmatic alternative provided by
(\ref{blackadsstring}). This solution describes 
a $(D-1)-$dimensional planar
adS black hole solution embedded in $D-$dimensional adS space. The adS
curvature of the $(D-1)-$dimensional slicing is the same as the global adS
curvature. It is easy to generalise this solution to
\be
\label{adsblackstring}
ds^2 = (\cosh(kz))^2\left[
V
dt^2 -{d\xi^2\over V }- \xi^2 dx_{(D-3),\kappa}^2\right]-dz^2
\ee
where $V(\xi)=\kappa+k^2\xi^2-{M\over \xi^{D-3}}$, $\kappa=0,\pm1$. 

To better understand the geometry, let $\kappa = M = 0$, and
consider the coordinate transformation
\be
ku = \xi^{-1} \sech kz \qquad , \qquad kv = \xi^{-1} \tanh kz
\ee
This transforms (\ref{blackadsstring}) into conformal
planar adS coordinates:
\be
ds^2 = {1\over k^2u^2} \left [ dt^2-dv^2-du^2-dx_n^2\right]
\ee
Thus the adS horizon $u\to\infty$ corresponds to $\xi\to0$,
and the adS boundary $u\to0$ to $|z|\to\infty$ (see figure 
\ref{fig:bhole}). Lines of constant $z$ are radial lines in the
$(u,v)$ plane, and constant $\xi$ are semi-circles centered on 
$u=v=0$. An adS braneworld corresponds to a line at constant $z$.
As pointed out in \cite{EHM2,KMP}, it is possible for two positive
tension adS braneworlds to localize gravity, which would correspond 
to two contant $z$-trajectories, one with $z>0$ and the other $z<0$.
\FIGURE{
\includegraphics[height=15cm]{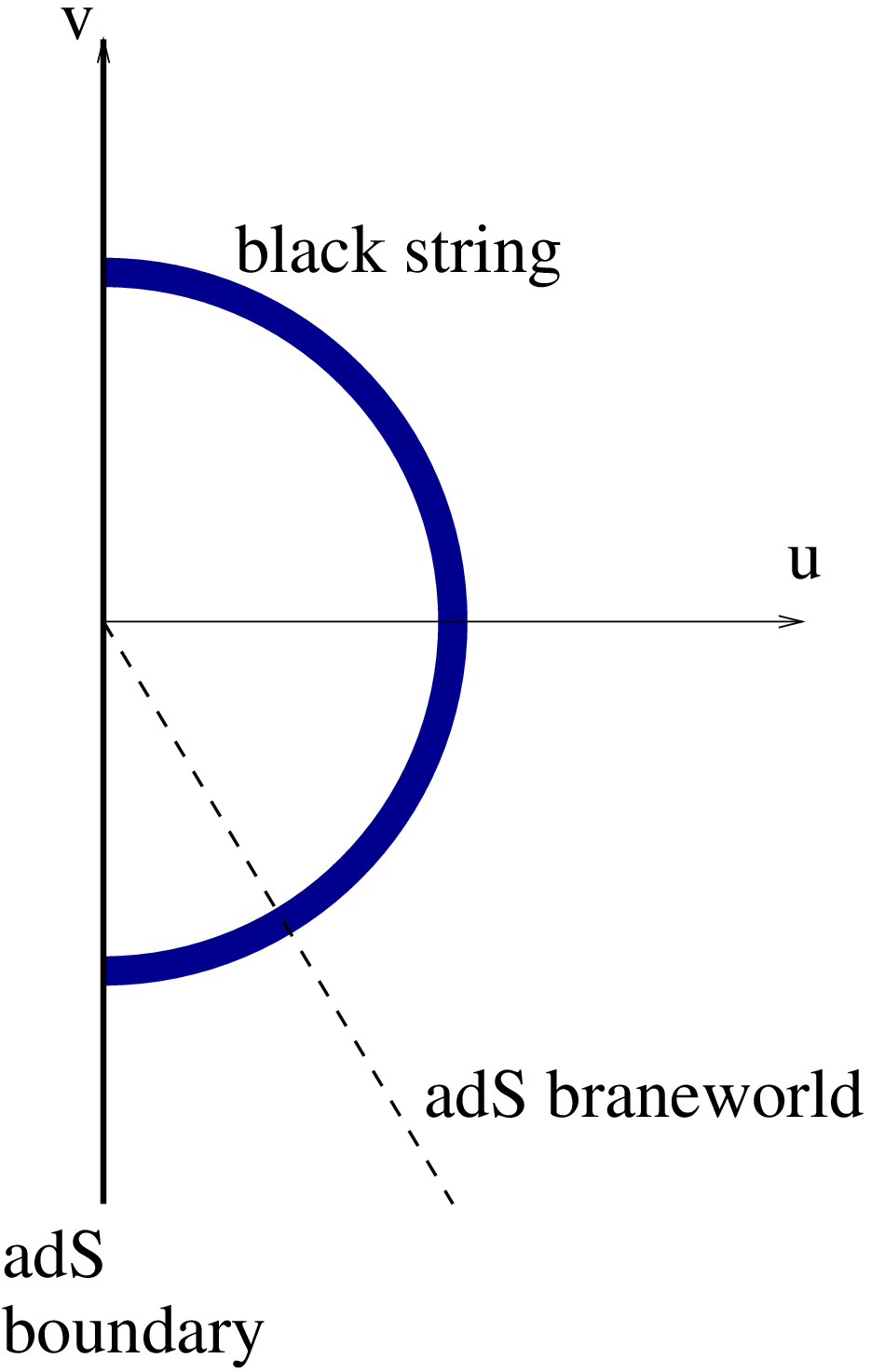}
\caption{The braneworld black hole spacetime in conformal coordinates.
The horizon is at $(u^2+v^2) = k^{-2} M^{-2/(n+1)}$, and the physical
spacetime exterior to the horizon corresponds to the {\it interior}
of the semi-circle.}
\label{fig:bhole}
}

Now consider $M\neq0$. Then since $\xi=$constant corresponds to
a semi-circle centered on $u=v=0$, the black brane solution
(\ref{blackadsstring}) corresponds to an horizon at fixed
($u^2+v^2$). In other words, the adS conformal plane has now 
been truncated by the horizon at fixed radius, and in particular,
the horizon of the black hole extends out all the way to the
adS boundary.
The former adS horizon now corresponds to the location of the
singularity and is cloaked by the black hole event horizon.
Since the adS plane is truncated before the adS horizon is reached,
the physical spacetime is well-behaved. Therefore, at the
price of introducing an adS braneworld, we can have a 
black hole on the brane which may well be stable, or at least
stable for a sufficiently wide parameter range to be useful.

Clearly the properties of such black holes are interesting, particularly
the issue of the holographic interpretation of such a truncated
spacetime, with an horizon on the adS boundary, and
will be the topic of a future study.

\section{Conclusions and further questions}

In this paper we have found and studied extensively solutions to the Einstein
equations in arbitrary dimensions under the assumption of axial 
symmetry and staticity.  We analysed the field equations, finding 
a duality relation which permitted us to map between solutions of
different dimensionality and geometrical characteristics. The duality
relation maps vacuum solutions with `internal' curvature, \ie\ where
there is a nonabelian SO($n$) symmetry group of the spacetime,
to adS spacetimes with no
internal curvature by quite simply associating the unique curvature
(or length) scale of one solution to the other. For example we
saw that the usual black brane solution with cylindrical horizon
was dual to the planar adS black
hole. 

Three classes of exact solutions were found in both 
vacuum and in the presence of a bulk cosmological constant. 
These solutions are related to lower dimensional spacetimes
with a dilaton. The first class depend on only one variable
and represent the most general solutions of this type. They
represent an axisymmetric generalization of the cosmic $p$-brane
solution of \cite{CPB}. The second class of solutions depend
on both variables, although the dependence on one of these is
linear. The solutions were not all given explicitly, rather,
their existence demonstrated and asymptotic properties derived from a
dynamical systems analysis of the equations of motion. There was 
however, a single stable critical point exact solution for the Thorne
vacuum $\kappa =1$ system, and the Weyl cosmological constant
system first discovered in \cite{Langlois:2001dy} in the context of
braneworld cosmology and later
discussed in \cite{christos}. Finally, the
most general two-dimensional solution with the simplest form of
$\phi$ was found to be either the cosmic $p$-brane/Rindler 
spacetime or one of a second family of spacetimes which have
a singular angular dependence in the metric. 

What has also become quite clear within our analysis is that  higher
dimensional spacetimes are richer in solutions, which is not
surprising, but also that they are generically more singular. 
Furthermore, solutions which seem more physically acceptable such 
as the black brane solutions of Horowitz and Strominger \cite{HS}
are unstable to small perturbations \cite{GL} and therefore
should be considered as unphysical.
Should one then question cosmic censorship in higher than four dimensions?
Or simply non-supersymmetric gravity in higher dimensions? Or perhaps
Einstein gravity, rather than Einstein-Gauss-Bonnet gravity 
\cite{lovelock} (see for example \cite{GB} and references within 
in the context of braneworld cosmology) in more
than four dimensions? Since we have not found an exhaustive classification
of solutions we cannot answer this question, although it is possible
that we are simply looking for solutions with the wrong sets of
Killing symmetries. By this, we mean that it might be possible for
null singularities to be replaced by null horizons by a change in
the geometry of the internal spaces, just as the cosmic $p$-branes
become nonsingular as their worldbranes are allowed to inflate \cite{rinfl}.

\acknowledgments
It is a great pleasure to thank Brandon Carter,
Roberto Emparan, Bob Johnson and Toby Wiseman for
discussions. We would particularly like to thank Peter Bowcock,
David Langlois and Jihad Mourad for helpful comments, discussions 
and encouragement in the early stages of this work.
CC is supported by the CNRS, and 
RG by the Royal Society.

\end{document}